# The Smart Shower

*An innovative device that saves water wastage during shower*

Umair Atique Syed
Dept. of Electrical Engineering
Institute of Applied Technology
Al-Ain, UAE
Umair.syed@iat.ac.ae

Uma Kandan Muniandy
Dept. of Electrical Engineering
Institute of Applied Technology
Al-Ain, UAE
Uma.muniandy @iat.ac.ae

*Abstract*—*The smart shower is an intelligent device that saves the water during the shower. It uses the indicator lamps that inform the user of the amount of the water. Like the traffic signal it has three sets of lamps, green, yellow and red, each indicating the amount of time spent. This device brain is the Siemens Logo PLC.*

## I. Introduction

We waste large volume of water while taking a shower every day. On average, we waste 7 gallons of water per minute (ref 1). So for a shower session of 10 minutes means 70 gallons of water is wasted. Whilst we take water and the shower for granted there are people where water is very precious like in UAE. UAE government is spending huge sums of money for desalinating of seawater so that the water can be supplied right in the taps and showers since ground water is not suitable for drinking and used mainly for agriculture.

The aim of this project is to reduce the water usage during a shower session. The smart shower uses indication system that informs the user about the amount of water they used. There are three set of lights, green, yellow and red; the green means that the user is within the water usage limit per session, the yellow light indicates that the user has almost reached the water usage limit and needs to finalize the process, while red means that the user has reached the limits and needs to go out as soon as possible.

The brain of this project is Siemens logo soft PLC (Programmable logic controller). The input is one ultrasonic sensor and outputs are 3 sets of lights and a water pump.

## II. Part Used

Following are the major electrical items has been used in the project:
- Ultrasonic Sensor – 24VDC, PNP, Range 10 cm
- Programmable Logic Controller – Siemens logo Soft
- Set of Red, Yellow and Green Lamps – 240VAC
- Power supply- 24VDC

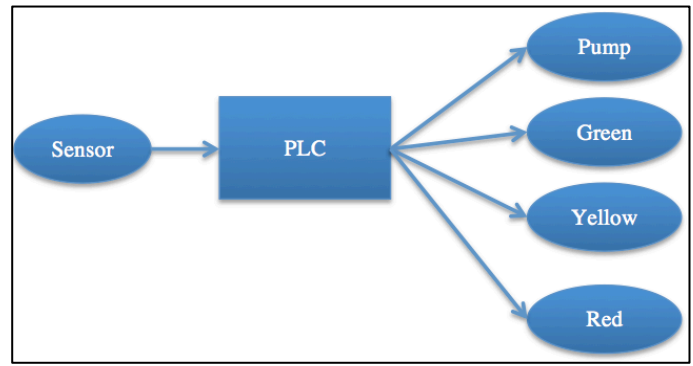

*Figure 1 The block diagram of the project*

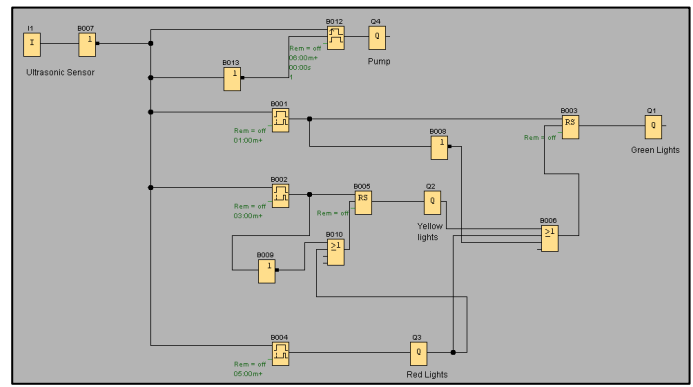

*Figure 2 PLC program*

## III. Working

The working depends on the five conditions of the user interference with the system.

### A. Condition 1

Condition 1 applies when user stands in the shower indefinitely. Following are the steps that are occurred.
 1. Ultrasonic sensor detects that the user in front of the shower.
 2. The pump starts pushing the water from the showerhead.
 3. Set of green lamps turns on for one minute (indicating the water usage is within the safe limits).

4. After one minute, the set of green lamps turns off and the set of yellow lamps turn on (indicating the water usage is about to reach the waste limits).
5. After 3 minutes, the set of yellow lamps turns off and the set of red lamps turn on (indicating the water usage has reached the waste limit and the user needs to finish shower).
6. After 5 minutes, the set of red lamps turns off.
7. The pump stops working 1 minute after the red light is turned off.

### B. Condition 2

Condition 2 applied when the user stands in the shower until the red light turns on. Following are the step that are occurred.
1. Ultrasonic sensor detects that the user in front of the shower.
2. The pump starts pushing the water from the showerhead.
3. Set of green lamps turn on for one minute (indicating the water usage is within the safe limits).
4. After one minute, the set of green lamps turns off and the set of yellow lamps turn on (indicating the water usage is about to reach the waste limits).
5. After 3 minutes, the set of yellow lamps turns off and the set of red lamps turn on (indicating the water usage has reached the waste limit and the user needs to finish shower).
6. After 5 minutes, the set of red lamps turns off and once the user move, the pump stops.
  *

### C. Condition 3

Condition 3 applies when the user is still not done even after the water shutdown. Following steps are occurred.
1. Ultrasonic sensor detects that the user in front of the shower.
2. The pump starts pushing the water from the showerhead.
3. Set of green lamps turns on for one minute (indicating the water usage is within the safe limits).
4. After one minute, the set of green lamps turns off and the set of yellow lamps turn on (indicating the water usage is about to reach the waste limits).
5. After 3 minutes, the set of yellow lamps turns off and the set of red lamps turn on (indicating the water usage has reached the waste limit and the user needs to finish shower).
6. After 5 minutes, the set of red lamps turns off.
7. The pump stops working 1 minute after the red light is turned off.
8. If the user needs to still use it, all he is suppose to do it move from the sensor and stand again in front of it and the cycle starts again.

### D. Condition 4

Condition applies when the user finished early or he/she needs to go out early. In this condition the cycle starts as usual and as soon as the user moves away from the shower, the cycle stops regardless of which stage of the cycle it is. The cycle can start again once the user gets back to the shower.

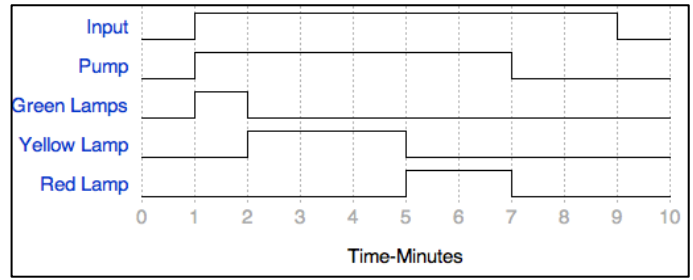

*Figure 3 The timing diagram*

## IV. FUTURE PLAN

Even though the project resolves a major daily problem but there is still a need for improvement in the project. We are considering following improvements for the project.

### A. Time Display

The project can be upgraded by putting the timer that displays the time taken for the shower session.

### B. Water Usage Display

The hall effect water flow sensor can be used to display the volume of water used in each session.

### C. Inbuilt Lamps

The plan is to construct a shower head with the lamps so that color of the water changes.

### D. Solar power

Solar power is to be included so that the system does not need any need of external electricity supply.

### E. Solar power

We plan to redesign the whole prototype to make it more compact.

## REFERENCES

[1]   *SWISH team, Community Science Action Guide,*

   http://fi.edu/guide/schutte/howmuch.html

[2]   Kevin Collins (2007). *PLC Programming for Industrial Automation.*

   Exposure Publisher. Reprint.

[3]   Stephen P. Tubbs (2005), *Programmable Logic Controller (PLC) Tutorial: Circuits and Programs for Rockwell Allen-Bradley MicroLogix and SLC 500 Programmable Controllers for Electrical Engineers and Technicians.* Stephen Philips Tubbs Publisher.

[4]   Jon Clift, Amanda Cuthbert (2007). *Water: Use Less-Save More: 100 Water-Saving Tips for the Home.* Chelsea Green Publishing